\documentclass[reprint,prb,aps,amsmath,amssymb,floatfix,superscriptaddress,nofootinbib,longbibliography]{revtex4-2}

\usepackage{physics}
\usepackage{bm}
\usepackage{graphicx}
\usepackage{xcolor}
\usepackage[english]{babel}
\usepackage{amsfonts}
\usepackage{amssymb}
\usepackage[colorlinks = true,
            linkcolor = red,
            urlcolor  = blue,
            citecolor = magenta,
            anchorcolor = red]{hyperref}
\usepackage{amsmath}
\def\be{\begin{equation}}
\def\ee{\end{equation}}

\usepackage{braket}
\usepackage[normalem]{ulem}

\newcommand{\beq}{\begin{equation}}
\newcommand{\eeq}{\end{equation}}

\newcommand{\gL}{\gamma_\mathrm{L}}  
\newcommand{\Ld}{\mathrm{L}}  
\newcommand{\gC}{\gamma_\mathrm{C}}  
\newcommand{\Cd}{\mathrm{C}}  
\newcommand{\gR}{\gamma_\mathrm{R}}  
\newcommand{\Rd}{\mathrm{R}}  
\newcommand{\Lb}{\mathsf{L}}  
\newcommand{\GL}{\Gamma_\mathsf{L}}  
\newcommand{\Rb}{\mathsf{R}}  
\newcommand{\GR}{\Gamma_\mathsf{R}}  
\newcommand{\IT}{I_\mathrm{T}}  

\begin{document}

\title{Collective charge measurement in quantum dot chains: controlling barrier occupation and tunneling current
}

\author{Alok Nath Singh}
\affiliation{Department of Physics and Astronomy, University of Rochester, Rochester, NY 14627, USA}
\affiliation{Institute for Quantum Studies, Chapman University, Orange, CA 92866, USA}

\author{Rafael S\'anchez}
\affiliation{Departamento de F\'isica Teorica de la Materia Condensada, Universidad Aut\'onoma de Madrid, 28049 Madrid, Spain\looseness=-1}
\affiliation{Condensed Matter Physics Center (IFIMAC), Universidad Aut\'onoma de Madrid, 28049 Madrid, Spain}
\affiliation{Instituto Nicol\'as Cabrera (INC), Universidad Aut\'onoma de Madrid, 28049 Madrid, Spain }

\author{Andrew N. Jordan}
\affiliation{Institute for Quantum Studies, Chapman University, Orange, CA 92866, USA}
\affiliation{Department of Physics and Astronomy, University of Rochester, Rochester, NY 14627, USA}
\affiliation{The Kennedy Chair in Physics, Chapman University, Orange, CA 92866, USA}
          
\begin{abstract}

We investigate nonequilibrium transport in a triple-quantum-dot (TQD) system, where the central dot acts as a discrete tunnel barrier, subject to continuous monitoring by a quantum point contact (QPC) that is capacitively coupled to all three dots with independently tunable strengths. 
We show that this global measurement scheme affects transport in a qualitatively distinct manner from single-site measurement. By engineering structured dephasing, measurement provides a significant improvement in the barrier occupation and tunneling current. In the strong-measurement limit, the steady state becomes independent of the underlying Hamiltonian parameters, and the barrier occupation can approach $1/2$ for suitable measurement configurations. {We identify an optimal measurement configuration that maximizes the steady-state current and show that near-optimal performance can be achieved with a simple central-dot readout scheme.}

\end{abstract}

\maketitle


\section{Introduction}
Continuous quantum measurement plays a dual role in open quantum systems: the measurement record conditions individual stochastic trajectories, while measurement backaction affects the ensemble-averaged dynamics through dephasing and diffusion in state space \cite{andrew_book}. In mesoscopic conductors, the capacitive coupling to a quantum point contact (QPC)~\cite{vanhouten_qpc_1992} provides a particularly natural and experimentally tunable platform for weak continuous detection of charge configurations in few electron quantum dots~\cite{field_measurements_1993,elzerman_few_2003}. It allows for monitoring the statistics of transport~\cite{schleser_time_2004,fujisawa:2006,gustavsson:2006} and gives access to which-path detection in interferometers~\cite{buks_dephasing_1998}, the state of spin qubits via charge-to-spin conversion~\cite{elzerman_single_2004} or thermodynamic quantities~\cite{kung_irreversibility_2012,saira_test_2012,koski:2014,koski2014_experimental_2014,hofmann_measuring_2016,hofmann_equilibrium_2016,hofmann_heat_2017,chida_seebeck_2022,chida_coulomb_2025}. In quantum dot arrays, a single detector can be used to monitor the internal dynamics~\cite{kung_irreversibility_2012,braakman_long_2013,chida_coulomb_2025}. Being noninvasive, the detector has nevertheless a backaction on the system dynamics that can be described within stochastic master equation frameworks \cite{gurvitz1997measurements, clerk2010introduction, brandes2005coherent,zilberberg2014}.

\begin{figure}[b]
    \centering
    \includegraphics[width=1\linewidth]{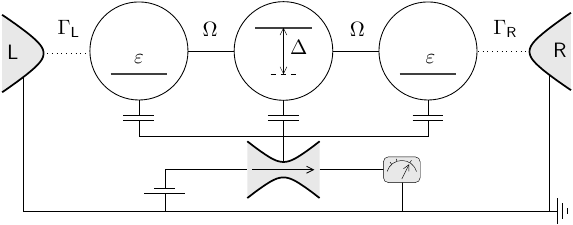}
\caption{Scheme of the triple quantum dot coupled to two terminals ($l={\sf L,R}$) via tunneling rates $\gL$. The energies of the singly-occupied quantum dots are represented: the central dot is split by $\Delta$ with respect to the other two, at $\varepsilon$. The nearest neighbor hopping is $\Omega$. The QPC is capacitively coupled to all three dots with measurement strengths $\gamma_\Ld,~\gamma_\Cd$, and $\gamma_\Rd$ respectively.
    }
    \label{fig:scheme}
\end{figure}

A triple quantum dot (TQD) forms the minimal setup to define the tunneling of single electrons with a particular energy $\varepsilon$ through a barrier defined by the center quantum dot being detuned by $\Delta$~\cite{granger_three_2010,hsieh_physics_2012}, as sketched in Fig.~\ref{fig:scheme}. The direct tunneling between the two extremal dots has been observed in the form of narrow transport resonances~\cite{busl_bipolar_2013,sanchez_longrange_2014,tormoQueralt_novel_2022} and via charge detection~\cite{braakman_long_2013} and been used to operate two-qubit gates~\cite{baart_coherent_2017}.
Continuous monitoring of the central dot reveals that measurement backaction can substantially increase its occupation~\cite{singh2025capturing} and, over an intermediate regime, enhance transport before strong monitoring suppresses current through a quantum-Zeno mechanism \cite{facchi2008quantum, greenfield2025unified}. Those results demonstrated that measurement does not merely reveal tunneling dynamics but actively modifies them. In that setting, the effect could be interpreted as the measurement partially localizing the electron in the virtual state (over the barrier), thereby modifying the tunneling processes. {This effect is fundamentally different from other types of backaction of environmental degrees of freedom on the tunneling process which treat the barrier as the separation of two allowed regions~\cite{brandes_adiabatic_2002,hussein:2015}, i.e., with no internal states.}

The present work builds directly on the framework developed in Ref.~\onlinecite{singh2025capturing}, going beyond single-site monitoring by considering a global measurement scheme in which a single QPC simultaneously monitors multiple dot occupations with independently tunable strengths, as depicted in Fig.~\ref{fig:scheme}. This enlarged control space allows us to engineer spatially structured dephasing across the device.
Here, the intuitive picture of measurement simply “localizing” the electron in the monitored dot is no longer sufficient. When several sites are monitored, it is not the absolute magnitude of a single measurement rate that determines the steady state, but rather the pattern of dephasing they induce between different site pairs. By selectively suppressing coherences across chosen bonds, the measurement can control the transport across the device.

A central result of this work is that global monitoring has a considerably different impact on transport through the TQD compared to that of single-site detection. In particular, it can produce stronger measurement-induced enhancements of both virtual-state occupation and measurement-assisted tunneling. The interplay between coherence and spatially structured dephasing therefore gives rise to regimes that are not accessible when only a single dot is monitored.
We analyze the ensemble-averaged long-time steady states obtained from the stochastic master equation and derive compact analytic expressions in the strong-measurement limit.
In this limit, the dynamics become independent of the system parameters and the virtual state occupation can rise up to 1/2 with appropriate measurements. {Importantly, we identify the measurement configuration that maximizes the steady-state current and show that near-optimal performance can be achieved with simpler measurement schemes, in particular via central-dot readout with appropriately tuned strength.}

The remainder of the text is organized as follows. The model of the TQD-QPC system and the diffusive detection regime are presented in Secs.~\ref{sec:model} and \ref{sec:diffusive} respectively. Results for the steady state transport using strong dephasing are discussed in Secs.~\ref{sec:steadyst}, with conclusions remarked in Sec.~\ref{sec:conclusion}.

\section{Model}
\label{sec:model}

We consider a TQD composed of three sites labeled L, C, and R, see Fig.~\ref{fig:scheme}. The energies of the different quantum dots can be tuned via plunger gates~\cite{granger_three_2010}. We fix a configuration where the left (L) and right (R) dots have identical on-site energies $\epsilon$ and are each coupled to separate thermal reservoirs, denoted by $\Lb$ and $\Rb$. The central dot (C) is detuned in energy by an amount $\Delta$ relative to L and R. Coherent tunneling occurs between L–C and C–R with identical hopping amplitude $\Omega$.

Strong on-site and inter-dot Coulomb repulsion  ensures that at most a single electron can be present inside the TQD system (the Coulomb blockade regime). In this regime the relevant dynamics are captured by a spinless, interaction-free Hamiltonian,
\begin{equation}\label{eq:Ham}
    \hat{H}_{\mathrm{TQD}} = \sum_i \varepsilon_i \hat{c}_i^\dagger \hat{c}_i 
- \sum_{i \ne j} \left( \Omega \hat{c}_i^\dagger \hat{c}_j + \text{H.c.} \right),
\end{equation}
where $\hat{c}_i$ annihilates an electron on dot $i \in \{{\rm L, C, R}\}$ with on-site energy $\varepsilon_i$, and $\hat n_i = \hat c_i^\dagger \hat c_i$ is the corresponding number operator.
In the regime $\Delta \gg \Omega$, hybridization of the central dot with the outer dots is strongly suppressed. Transport between L and R then proceeds predominantly via virtual occupation of the detuned central level. In this limit the TQD acts as a discrete tunnel barrier, with $\Delta$ controlling the effective barrier height. A standard second-order perturbative treatment yields an effective coherent coupling $\Omega_{\rm eff} = \Omega^2 / \Delta$~\cite{ratner_bridge_1990}. The characteristic response time of the detector is taken to be much smaller than the intrinsic virtual-state lifetime, $\hbar/\Delta$, in order to be sensitive to its transient occupation.

The reservoirs are modeled as noninteracting fermionic continua described by
$\hat H_{\rm res} = \sum_{l}\sum_k \epsilon_{k,l}\hat{d}_{k,l}^\dagger \hat d_{k,l}$,
where $\hat{d}_{k,l}$ annihilates an electron of energy $\epsilon_{k,l}$ in reservoir $l \in \{{\rm \Lb, \Rb}\}$. The tunneling between the TQD and the reservoirs is given by
$
\hat{H}_{\rm tun} = \sum_{l,k}\tau_{l}\hat{d}^{\dagger}_{k,l}
\hat{c}_{l}+\text{H.c.},
$
where $\tau_l$ sets the (weak) coupling strength to reservoir $l$.

\section{Diffusive Quantum Measurement}
\label{sec:diffusive}

We consider weak continuous measurement by the QPC, which is capacitively coupled to all three dots, see Fig.~\ref{fig:scheme}. Inside the QPC, the transmittance of the saddle point constriction depends on whether an electron is present in either of the dots~(with transmittance ${\cal T}_i$ for dot $i$) or not~(with transmittance ${\cal T}_0$). In the diffusive quantum measurement limit, transmittance is not affected significantly by the occupation, i.e., $|{\cal T}_i-{\cal T}_0|\ll {\cal T}~\forall~i \in \{\rm L,C,R\}$ \cite{andrew_book}, where ${\cal T} = ({\cal T}_0+\Sigma_i {\cal T}_i)/4$ is the average transmittance (see Appendix~\ref{app:diffusive_condition} for more details).

Assuming weak coupling to the reservoirs, we use the stochastic master equation, in It\^o form~\cite{andrew_book}, to evaluate the TQD state evolution,
\beq\label{eq:ME}
\frac{d\hat\rho}{dt}=-\frac{i}{\hbar}[\hat{H}_{\text{TQD}},\hat\rho]+\sum_{l}({\cal L}_{\Gamma,l+}\hat\rho+{\cal L}_{\Gamma,l-}\hat\rho)+{\cal L}_\gamma\hat\rho,
\eeq
where 
${\cal L}_{\Gamma,\lambda}=\hat{L}_\lambda^{}\hat\rho\hat{L}_\lambda^{\dagger}-\frac{1}{2}\left(\hat{L}_\lambda^{\dagger}\hat{L}_\lambda^{}\hat\rho+\hat\rho\hat{L}_\lambda^{\dagger}\hat{L}_\lambda^{}\right)$
are the usual Lindblad superoperators corresponding to the reservoirs, with the jump operators related to tunneling out, $\hat{L}_{l+}=\sqrt{W_{l+}}|l\rangle\langle0|$, or tunneling in, $\hat{L}_{l-}=\sqrt{W_{l-}}|0\rangle\langle l|$, of thermal bath $l={\sf L,R}$.
The tunneling rates are given by $W_{l+}=\Gamma_l f(\varepsilon_l-\mu_l)$ and $W_{l-}=\Gamma_l[1-f(\varepsilon_l-\mu_l)]$, where $f(E)=[1+\exp{(E/k_{\rm B}T)}]^{-1}$ is the Fermi function and $\mu_l$ is the electrochemical potential of reservoir $l$ being at temperature $T$. The rate between dot $l$ and reservoir $l$ is $\Gamma_l=2\pi\hbar^{-1}|\tau_l|^2\nu_l$, with $\nu_l$ being the density of states in lead $l$. The stochastic measurement superoperator is given by
\beq \label{eq:measurement_superoperator}
\begin{aligned}
{\cal L}_\gamma \hat{\rho}=~&\hat{L}_\gamma^{}\hat\rho\hat{L}_\gamma^{\dagger}-\frac{1}{2}\left(\hat{L}_\gamma^{\dagger}\hat{L}_\gamma^{}\hat\rho+\hat\rho\hat{L}_\gamma^{\dagger}\hat{L}_\gamma^{}\right)\\
&+ (\hat{L}_\gamma\hat\rho + \hat\rho\hat{L}_\gamma^{\dagger} - \langle \hat{L}_\gamma + \hat{L}_\gamma^\dagger \rangle \hat\rho) \frac{dW}{dt}
\end{aligned}
\eeq
where the jump operator is 
\be\label{eq:meas_jump_op}
\hat{L}_\gamma = \sqrt{\gamma_\Ld}|L\rangle \langle L| + \sqrt{\gamma_\Cd}|C\rangle \langle C| + \sqrt{\gamma_\Rd}|R\rangle \langle R|,
\ee 
and $dW$ is the Wiener increment associated with each measurement readout, being a zero mean random variable and obeying $dW^2=dt$ ($dt$ being the time interval). The measurement strength for dot $i$ is given by $\gamma_i=eV({\cal T}_0-{\cal T}_i)^2/2h{\cal T}(1-{\cal T})$~\cite{andrew_book}, where $V$ is the voltage applied across the QPC and $h$ is Planck's constant.
A derivation of this expression is provided in Appendix~\ref{app:measurement_strength}. Note that ${\cal L}_\gamma\rho$ induces crossed terms which make the evolution of the coherences, $\dot\rho_{jk}$, depend on the detector as $\langle j|{\cal L_{\gamma}}\hat\rho|k\rangle=-D_{jk}\rho_{jk}$, with $D_{jk}=(\sqrt{\gamma_j}-\sqrt{\gamma_k})^2/2$ denoting the dephasing rate induced between the dots $j$ and $k$. This is different from the case where the system charge is monitored by three different detectors and the dephasing rate is additive, given by $(\gamma_j+\gamma_k)/2$~\cite{contrerasPulido_dephasing_2014}. 


We consider the situation where the reservoirs are at the same temperature, but have a large voltage bias across them, such that $f(\varepsilon_{\rm L}~-~\mu_{\rm L})~\approx~1$ and $f(\varepsilon_{\rm R}~-~\mu_{\rm R})~\approx~0$. The local master equation is a valid approximation in this limit when $\Omega \ll \Gamma_{\Lb},\Gamma_{\Rb}$ \cite{gurvitz_microscopic_1996}. {In GaAs double quantum dots, it has been demonstrated that  $\Omega$ can be tuned to a few $\mu\text{eV}$, while $\GL$ and $\GR$ can reach tens of $\mu\text{eV}$ \cite{hayashi2003coherent}, allowing access to this regime. The detuning $\Delta$ can also be independently tuned to be orders-of-magnitude higher than $\Omega$ \cite{petta2005coherent}}. Note that because of this large bias, our results here do not depend on $\varepsilon$ nor on bath temperatures. We use Eq.~(\ref{eq:ME}) to find a coupled differential equation for each element of $\hat{\rho}$ (see Appendix~\ref{app:SME} for complete expressions of the master equations), which are then solved numerically to get the state evolution. The particle current is then simply given by $I_{\rm T}=\GR\rho_{\rm RR}$.

The stochastic master equation in Eq.~(\ref{eq:ME}) describes the conditional evolution of the TQD state under continuous monitoring. Each realization of the Wiener increment $dW$ corresponds to a particular measurement record and therefore generates a distinct quantum trajectory. While the deterministic part of the evolution captures coherent tunneling and dissipative coupling to the reservoirs and the detector, the stochastic term encodes the measurement backaction associated with the QPC current fluctuations.
In the diffusive limit, the measurement signal is weak and noisy, such that individual trajectories exhibit pronounced fluctuations in both populations and coherences. However, averaging over many realizations eliminates the stochastic contribution, effectively setting $dW \rightarrow 0$. The resulting ensemble-averaged dynamics are governed solely by the deterministic part of the master equation and describe the unconditional evolution of the system.

\begin{figure}[t]
    \centering
    \includegraphics[width=1\linewidth]{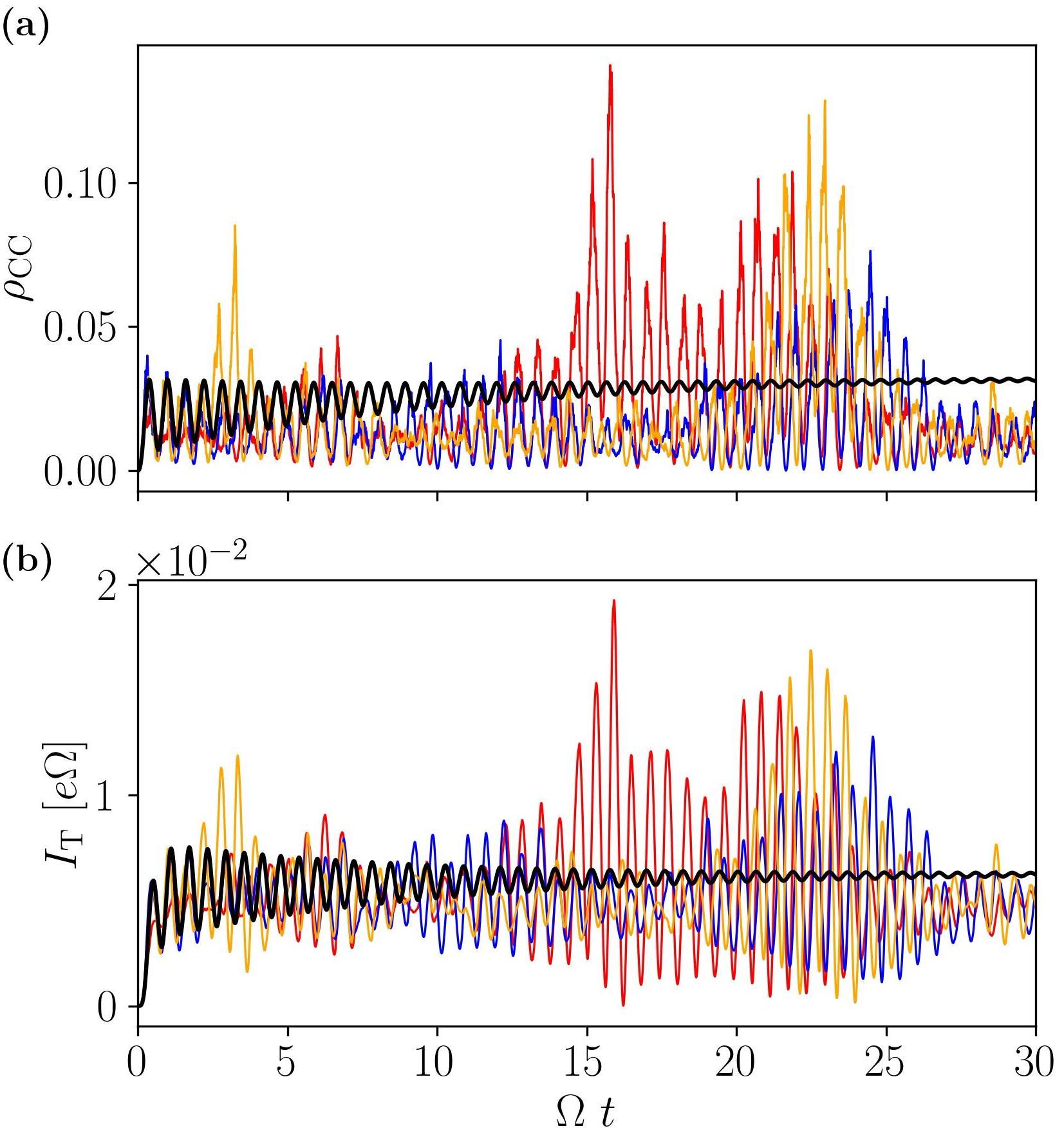}
\caption{Central dot occupation, $\rho_{\rm CC}$ (a), and TQD current, $I_{\rm T}$ (b), plotted as a function of time. The initial state is $\ket{\rm 0}\bra{\rm 0}$. The black curve correspond to the ensemble-averaged evolution, and the colored curves are individual stochastic trajectories. For the latter, time averaging is done over a rectangular window of $0.1~\Omega^{-1}$. Parameters: $\Gamma_{\Lb} = 10~\Omega$, $\Gamma_\Rb = 8~\Omega$, $\Delta=10~\Omega$, $\gamma_\Ld = \Omega$, $\gamma_\Cd = 2~\Omega$, and $\gamma_\Rd = \Omega$.
    }
    \label{fig:stoch_traj}
\end{figure}

Figure~\ref{fig:stoch_traj} illustrates this distinction. The colored curves represent individual stochastic trajectories of the central-dot occupation, $\rho_{\rm CC}$, and the current, $\IT$, passing from bath $\Lb$ to $\Rb$ through the TQD. The black curves correspond to the ensemble average. Although individual trajectories may display temporary revivals or rapid decay of coherences depending on the measurement outcomes, the averaged dynamics exhibit smooth damping toward a unique nonequilibrium steady state.

\section{ Steady-State Barrier Occupation and Tunneling Current}
\label{sec:steadyst}
In what follows, we focus on the long-time steady state of the ensemble-averaged dynamics. In particular, we study how steady-state $\rho_{\rm CC}$ and $I_{\rm T}$ varies as the measurement strengths of the different dots are varied. The steady-state density-matrix elements are obtained by setting the Wiener increment $dW$ and all time derivatives to zero in the master equations, yielding the following system of linear equations:
\begin{gather}
\begin{aligned}
\label{eq:stationarymastereq}
0&=\Gamma_{\sf L}\,\rho_{00}-2\Omega\,\Im\rho_{\rm LC}\\
0&=\Im\rho_{\rm LC}+\Im\rho_{\rm RC}\\
0&=\Gamma_{\sf R}\,\rho_{\rm RR} +2\Omega\,\Im\rho_{\rm RC}\\
0&=\Im(z_{\rm L}^*\rho_{\rm LC})
+ \Omega\,\Im\rho_{\rm LR}\\
0&=\Im(z_{\rm R}^*\rho_{\rm RC})
- \Omega\,\Im\rho_{\rm LR}
\\
0&=\Omega(\rho_{\rm CC}-\rho_{\rm LL})
- \Re(z_{\rm L}^{}\rho_{\rm LC})
\\
0&=\Omega(\rho_{\rm CC}-\rho_{\rm RR})
- \Re(z_{\rm R}^{}\rho_{\rm RC})\\
0&=\Omega\big(\Re\rho_{\rm RC}-\Re\rho_{\rm LC}\big)
+ \left(\frac{\Gamma_{\sf R}}{2}
+ D_{\rm LR}\right)\Im\rho_{\rm LR}
\end{aligned}
\end{gather}
as well as $\Re\rho_{\rm LR}=0$ and $\sum_{i}\rho_{ii}=1$. Here we have defined $z_{\rm L}=\Delta-iD_{\rm LC}$ and $z_{\rm R}=\Delta-i(D_{\rm RC}+\gR/2)$, see Appendix~\ref{app:SME} for more details.
From these equations we get $\Gamma_{\sf R}\rho_{\rm RR}=\Gamma_{\sf L}\rho_{00}$ and 
\be
\label{eq:pCCst}
\rho_{\rm CC}=\frac{1}{2}\left[\rho_{\rm LL}+\rho_{\rm RR}+\frac{1}{\Omega}
{\rm Re}(z_{\rm L}\rho_{\rm LC}+z_{\rm R}\rho_{\rm RC})\right],
\eeq
which relates the occupation of the virtual state to the coherences. {It is left-right asymmetric due to the non-equilibrium situation: In the high bias regime, tunneling to the right reservoir introduces decoherence, but tunneling from the left does not. In the equations, this is reflected in the imaginary parts of $z_{\rm R}$ and $z_{\rm L}$, respectively: while $z_{\rm R}$ depends on the coupling to the reservoirs, $z_{\rm L}$ does not.}










\begin{figure}[t]
    \includegraphics[width=\linewidth]{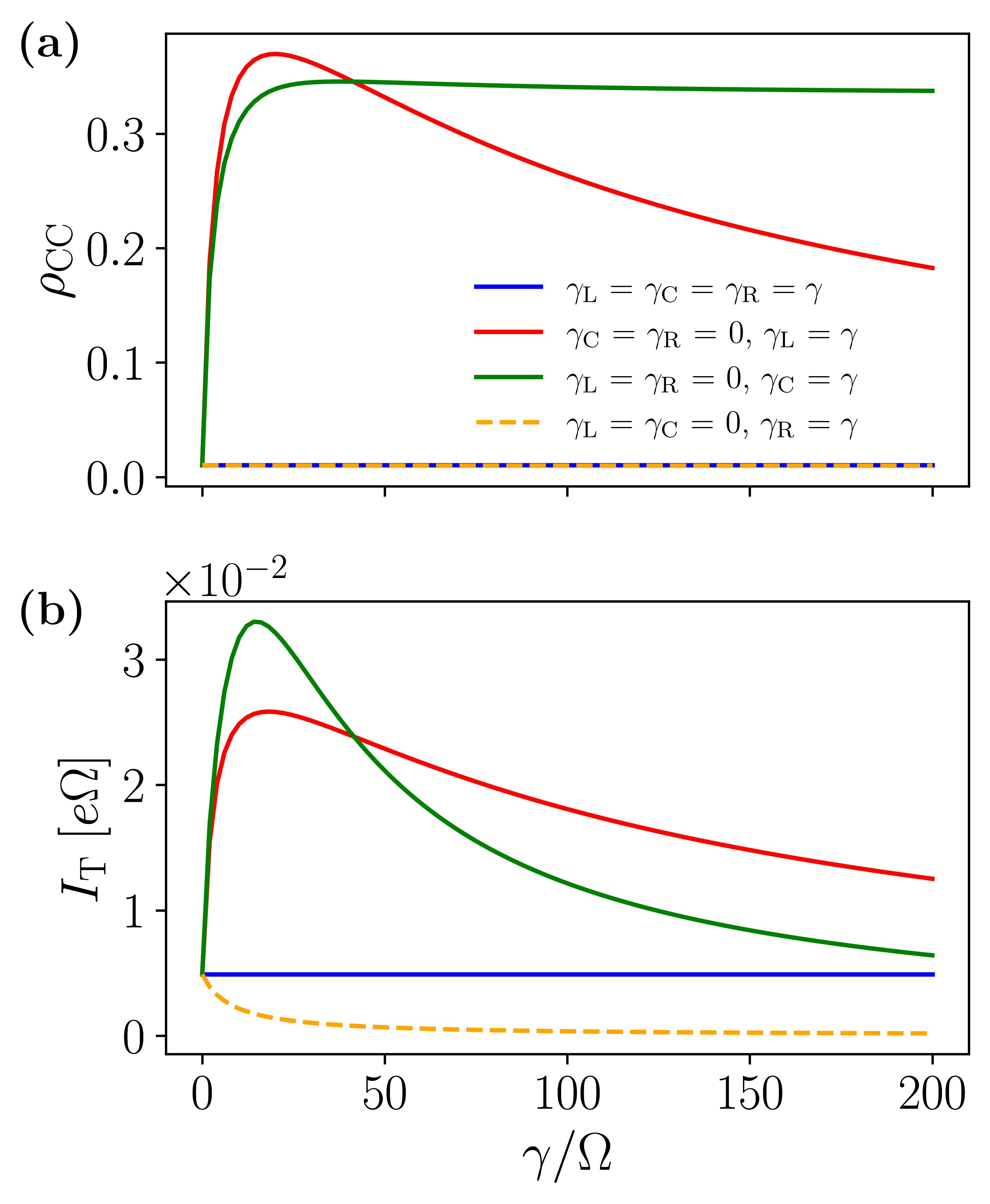}
    \caption{Steady-state central dot occupation (a) and TQD current (b) as a function of measurement strengths. Cases with different measurement-strength proportions, including uniform detection of all sites and local detection of individual sites, are plotted in different colors. Other parameters: $\Delta = 10\,\Omega$, $\Gamma_\Lb = 10\,\Omega$, and $\Gamma_\Rd = 8\,\Omega$.  
    }
    \label{fig:steady_state}
\end{figure}

As anticipated in Sec.~\ref{sec:diffusive}, the influence of measurement on the steady states enters in the coherences between two sites, $\rho_{jk}$, and is hence captured by the combination of dephasing rates $D_{jk}$ rather than by the individual rates $\gamma_j$ or $\gamma_k$ ($j,\,k \in \{{\rm L, C, R}\}$ and $D_{jk} = D_{kj}$). The interpretation is simple: a detector coupled similarly to dots $j$ and $k$ cannot resolve their occupation and hence does not decohere them~\cite{trigal2026noninvasive}. Considering three dots, the effect has, however, some less expected consequences, e.g., the dynamics of the TQD system are exactly the same whether $\gL:\gC:\gR=0:1:0$ or $\gL:\gC:\gR=1:0:1$ because the dephasing rates come out to be the same. In both cases, the detector can only distinguish whether an  electron in the TQD is in the center dot or not. {Another useful consequence is that the measurement configuration $\gL:\gC:\gR=a:g:a$ is identical (in terms of steady states) to $\gL:\gC:\gR=0:h:0$, where $g$ and a are arbitrary positive numbers, and $h=(\sqrt{g}-\sqrt{a})^2$. Thus, the idealized situation considered in Ref.~\onlinecite{singh2025capturing}—where the QPC is only coupled with the central dot—can also be realized in experiments where the outer dots are coupled with the QPC too, as long as they are both coupled equally strongly. A detailed discussion of equivalent measurement configurations is provided in Appendix~\ref{app:equiv_meas}.}

\subsection{Uniform Detection}

Let us first consider the special situation when all measurement strengths, $\gamma_i=\gamma$, are equal i.e., the detector is blind to which quantum dot is occupied. From the above equations it is clear to see that the measurement here has no impact whatsoever on the steady state. In this case, the measurement jump operator, $\mathcal{L}$, becomes $\sqrt{\gamma}(\openone-\ket{0}\bra{0})$. 
This measurement only affects the coherences between the $\ket{0}\bra{0}$ and the $(\openone - \ket{0}\bra{0})$ subspaces. The TQD Hamiltonian exists inside the $(\openone - \ket{0}\bra{0})$ subspace, and thus, these measurements do not affect the coherent evolution of the system to any degree on average. We show this with the blue curves in Fig.~\ref{fig:steady_state} for $\rho_{\rm CC}$ and $\IT$. Note that measurement, however, does have an effect on conditional evolution [see Appendix~\ref{app:equal_measurement} for more details].

{
\subsection{Local Detection}

In the limit where the detector couples to a single quantum dot, the measurement backaction strongly depends on which site is being monitored relative to the transport direction set by the bias (from left to right). This is illustrated in Fig.~\ref{fig:steady_state}, where we show the occupation of the center dot, $\rho_{\rm CC}$, and the current $\IT$ for different choices of the measured site. 
{The high bias condition, where the left dot is coupled to an electron source and the right dot to an electron sink, links the occupation of the empty and right-dot states, $\rho_{00}=\rho_{\rm RR}\Gamma_{\sf R}/\Gamma_{\sf L}$. The occupation of the left quantum dot can then be obtained from the conservation of probability, $\rho_{\rm LL}=1-\rho_{\rm CC}-(1+\Gamma_{\sf R}/\Gamma_{\sf L})\rho_{\rm RR}$.
Full analytical expressions of the steady-state occupations are given in Appendix~\ref{app:occupations}.}

A remarkable case is when the detector is only coupled to the central dot, which is energetically forbidden ($\Delta\gg\Omega$). This case, with $D_{\rm LC}=D_{\rm RC}=\gC/2$ and $D_{\rm LR}=0$, is investigated in detail in Ref.~\onlinecite{singh2025capturing}, exhibiting the saturation of the barrier occupation at strong measurement strengths, while the current displays measurement-assisted tunneling at intermediate strengths, leading to a pronounced maximum in $\IT$ (see green curves in Fig.~\ref{fig:steady_state}). 
{The Zeno effect at large $\gC$ leads to a suppression of the current as $I_{\rm T}\to4\Omega^2/3\gC$, with the charge being distributed between the left and center quantum dots,  $\rho_{\rm LL}\to2/3$ and  $\rho_{\rm CC}\to1/3$, independently of the barrier height.}

{The behavior is somewhat similar when the detector is coupled only to the left dot (red curves in Fig.~\ref{fig:steady_state}), such that $D_{\rm RC}=0$ and $D_{\rm LC}=D_{\rm LR}=\gL/2$. The occupation of the center dot increases with the measurement strength as $\rho_{\rm CC}=\gL/(\Gamma_{\sf R}+2\gL)$ 
to leading order in $\Omega/\Delta$, and decays in the very strong detection limit as $\rho_{\rm CC}\to[\Gamma_{\sf R}^2+4(\Delta^2+\Omega^2)]/\Gamma_{\sf R}\gL$, implying a maximal occupation for intermediate strengths. 
This comes at the expense of the occupation of the left dot, which decreases for small $\gL$ and $\Delta\gg\Omega$ as $\rho_{\rm LL}=1-\gL/\Gamma_{\sf R}$ and saturates to 1 in the limit $\gL\to\infty$. 
In the latter limit, the combination of the barrier and the Zeno effect hence localizes the electron in the left dot, with all other occupations vanishing. Nevertheless, for intermediate measurement strengths, partial suppression of coherent tunneling enhances transport, 
\beq
I_{\rm T}=\frac{\Gamma_{\sf R}\gL+4\Omega^2}{\Gamma_{\sf R}+2\gL}\frac{\Omega^2}{\Delta^2}+{\cal O}\left(\frac{\Omega^4}{\Delta^4}\right),
\eeq
resulting again in measurement-assisted tunneling as long as $\Gamma_{\sf R}^2>8\Omega^2$, see Fig.~\ref{fig:steady_state}(b).
}

{In contrast, when the detector is coupled to the right dot (dashed orange curves in Fig.~\ref{fig:steady_state}), the occupation of the central dot remains almost independent of $\gR$ and the tunneling current is suppressed for any finite measurement strength.  In this case (now $D_{\rm LC}=0$, $D_{\rm RC}=\gR/2$), solving  Eqs.~\eqref{eq:stationarymastereq} we get simple analytical expressions for the most general configuration, giving:  
\begin{gather}
\begin{aligned}
\rho_{\rm CC}&=\frac{\Omega^2\Gamma_{\sf L}[4\Omega^2+\Gamma_{\sf R}(\Gamma_{\sf R}+\gR)]}{\Gamma_{\sf L}\Gamma_{\sf R}(\Gamma_{\sf R}{+}\gR)(\Delta^2+2\Omega^2)+4\Omega^4(3\Gamma_{\sf L}{+}\Gamma_{\sf R})}\\
I_{\rm T}&=\frac{4\Omega^2\Gamma_{\sf L}\Gamma_{\sf R}}{\Gamma_{\sf L}\Gamma_{\sf R}(\Gamma_{\sf R}{+}\gR)(\Delta^2+2\Omega^2)+4\Omega^4(3\Gamma_{\sf L}{+}\Gamma_{\sf R})}.
\end{aligned}
\end{gather}
In the regime of interest, where $\Delta\gg\Omega$, the occupation of the barrier depends weakly on $\gR$:
\begin{equation}
\rho_{\rm CC}=\left(1+\frac{4\Omega^2}{\Gamma_{\sf R}(\Gamma_{\sf R}+\gR)}\right)\frac{\Omega^2}{\Delta^2}+{\cal O}\left(\frac{\Omega^4}{\Delta^4}\right),
\end{equation}
while the current, $I_{\rm T}=\Gamma_{\sf R}\rho_{\rm RR}=4\Omega^4/\Gamma_{\sf R}(\Gamma_{\sf R}+\gR)\Delta^2+{\cal O}(\Omega^4/\Delta^4)$, is monotonically suppressed when increasing the measurement strength. Additionally, as $\rho_{00}\propto\rho_{\rm RR}$, the probability of finding the TQD empty also decreases with $\gR$. In this case, the Zeno effect manifests differently, as the measured dot is directly coupled to an electron sink, preventing the electron to localize in the monitored quantum dot, ${\rm R}$. The occupation is then distributed between ${\rm L}$ and ${\rm C}$, with $\rho_{\rm CC}\to\Omega^2/(\Delta^2+2\Omega^2)$, $\rho_{\rm LL}=1-\rho_{\rm CC}$ and $\rho_{00},\rho_{\rm RR}\to0$ as $\gR\to\infty$.
}

\subsection{Strong Measurement Limit}
{We can make some further analytical understanding in the strong measurement limit by neglecting the contribution of $\rho_{\rm LR}$. This limit is defined by any measurement-induced dephasing rate dominating over the other system parameters. It is justified to ignore $\rho_{\rm LR}$ as it is the most sensitive to the effect of the detector on every site of the array, as we have confirmed numerically. The steady-state results can be written in compact closed-form expressions:} 
\begin{gather}
\begin{aligned}\label{eq:strong_meas_current}
    \IT=&\bigg[\frac{1}{\Gamma_{\sf L}}+\frac{3}{\Gamma_{\sf R}}+\frac{1}{\Omega^{2}}\!\left(\frac{\Delta^{2}}{\Gamma_{\sf R}+D_{\rm RC}}+\Gamma_{\sf R}+D_{\rm RC}\right)\\
    &+\frac{1}{\Omega^{2}}\!\left(\frac{\Delta^{2}}{2D_{\rm LC}}+\frac{D_{\rm LC}}{2}\right)\bigg]^{-1} , 
\end{aligned}
\end{gather}
and
\begin{equation}\label{eq:strong_meas_occupation}
   \rho_{\rm CC}
=\left[\frac{1}{\Gamma_{\sf R}}
+\frac{1}{2\Omega^{2}}\!\left(\frac{\Delta^{2}}{\Gamma_{\sf R}{+}D_{\rm RC}}+\Gamma_{\sf R}+D_{\rm RC}\right)\right]\, \IT.
\end{equation}
For fixed values of $\Gamma_{\sf L},\,\Gamma_{\sf R}$, and $\Delta$, using the above expression, we find that $\rho_{\rm CC}\to 1/2$ when $D_{\rm RC}\to\infty$ but $D_{\rm LC}$ stays finite, irrespective of the other system parameters. $\Im \rho_{\rm RC}\to0$, and for the system to achieve steady-state, $\Im \rho_{\rm LC}$ needs to equal $\Im \rho_{\rm RC}$, see Eq.~\eqref{eq:stationarymastereq}. LC dephasing rate is not strong enough to help satisfy that condition, thus, the occupation in L and C should be very close to stop the current flow. Because of the large bias, $\rho_{00}$ and $\rho_{\rm RR}$ are close to 0, therefore, $\rho_{\rm CC}\approx\rho_{\rm LL}\approx1/2$, despite having very different energies.
As we shall see in the next subsection, the intermediate-strength peak in $\rho_{\rm CC}$ (e.g., the red curve in Fig.~\ref{fig:steady_state}(a)) never goes above 1/2. Thus, we believe that this is the highest possible value that $\rho_{\rm CC}$ can achieve, for any measurement configuration.

When both dephasing rates $D_{\mathrm{LC}}$ and $D_{\mathrm{RC}}$ are taken to the projective limit ($\to \infty$), we obtain $\rho_{\mathrm{CC}}\to 1/3$, consistent with the results presented in Ref.~\onlinecite{singh2025capturing}. This follows from current continuity. In steady state the particle current must be equal between L–C and C–R, and in the strongly dephased limit the local current is generated by classical incoherent hopping \cite{gurvitz1997measurements, gurvitz_microscopic_1996,facchi2008quantum}, which is proportional to the population difference. This leads to the condition: $\rho_{\rm LL}-\rho_{\rm CC}=\rho_{\rm CC} - \rho_{\rm RR}$, and thus, we get $\rho_{\rm LL}\approx2/3$ and $\rho_{\rm CC}\approx1/3$ in the large-bias limit.


{
\subsection{Mixed Measurement Strengths}

\begin{figure}
    \centering
    \includegraphics[width=1.0\linewidth]{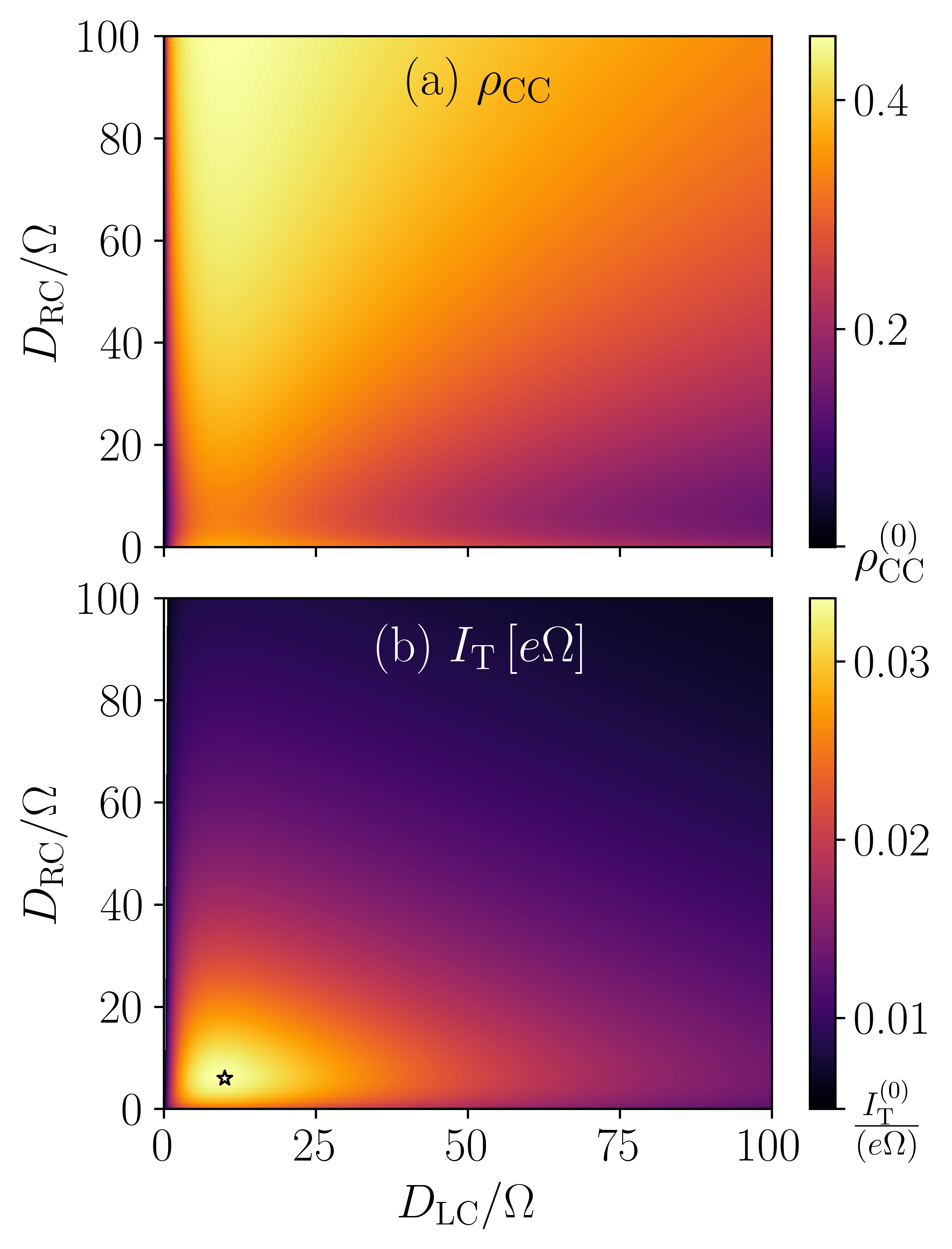}
    \caption{Steady-state central dot occupation (a) and TQD current (b) as a function of the dephasing terms $D_{\rm LR}$ and $D_{\rm RC}$. The star in panel~(b) denotes the point where the current is maximized. $\rho_{\rm CC}^{(0)}$ and $\IT^{(0)}$ denote the respective no-measurement values on the color bar. Other parameters: $\Delta = 10\,\Omega$, $\Gamma_\Lb = 10\,\Omega$, and $\Gamma_\Rd = 8\,\Omega$. 
    }
    \label{fig:ss_colormap}
\end{figure}

In this subsection, we analyze the global measurement scheme in more detail by exploring the full parameter regime. From the structure of the steady-state master equations, Eqs.~\eqref{eq:stationarymastereq}, it is clear that the dephasing terms $D_{\rm LC}$ and $D_{\rm RC}$ fully characterize the effect of measurement on the steady state, see also Appendix~\ref{app:equiv_meas}. This is because $D_{\rm LR} = (\sqrt{D_{\rm LC}} - \sqrt{D_{\rm RC}})^2$ is not an independent parameter, so that $(D_{\rm LC}$, $D_{\rm RC})$ span the entire physically accessible parameter space.

In Fig.~\ref{fig:ss_colormap}, $\rho_{\rm CC}$ and $\IT$ are plotted as a function of $D_{\rm LC}$ and $D_{\rm RC}$, sweeping a range from weak to strong measurements. In general, $\rho_{\rm CC}$ grows with $D_{\rm RC}$ and diminishes as $D_{\rm LC}$ is increased, except when $D_{\rm LC}\approx0$ and $\rho_{\rm CC}$ stays close to its low no-measurement value, as also shown in Fig.~\ref{fig:steady_state}. This plot verifies the observation from the preceding subsection that when $D_{\rm RC}\to\infty$ and $D_{\rm LC}$ is finite, $\rho_{\rm CC}\to1/2$. It is interesting that in the regime where $D_{\rm RC}$ is large, $\rho_{\rm CC}$ very rapidly rises from an extremely low value ($\approx0.01$ for the parameters chosen in Fig.~\ref{fig:ss_colormap}) to its maximum value around 1/2 with small increases of $D_{\rm LC}$. This rapid rise suggests a potential sensing application of any observable that can modulate $D_{\rm LC}$, in a similar fashion as how phase transitions are utilized for sensing purposes.

\begin{figure}[t]
    \includegraphics[width=\linewidth]{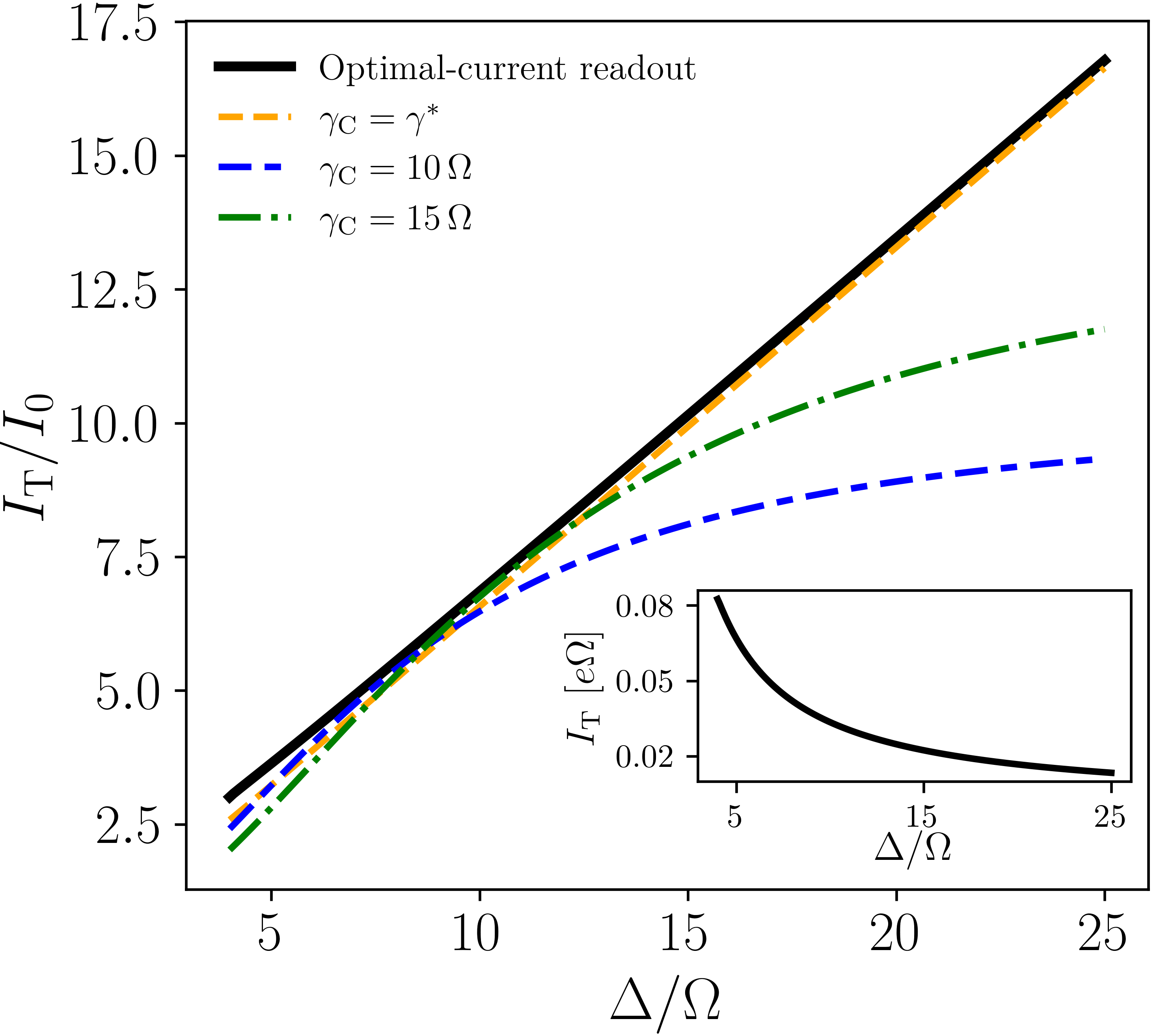}
    \caption{
    Steady-state current enhancement $I_{\rm T}/I_0$ versus detuning $\Delta$ for optimal-current readout described by Eq.~\eqref{eq:optimal_current}. It is compared with three simpler measurement configurations where only the central dot is measured ($\gL=\gR=0$): one with a variable strength of $\gamma^*=2\,\Delta$ (orange) and the other two with a fixed strength of $10\,\Omega$ (blue) and $15\,\Omega$ (green) The inset shows how the optimal current itself varies with $\Delta$.  
    Other parameters: $\Gamma_{\rm L}=10\,\Omega$, $\Gamma_{\rm R}=8\,\Omega$.
    }
    \label{fig:IT_over_I0}
\end{figure}

In regards to the tunneling current $\IT$ in Fig.~\ref{fig:ss_colormap}(b), we first observe measurement-assisted tunneling as either $D_{\rm RC}$ or $D_{\rm LC}$ are increased from 0, but when both dephasing terms reach $D_{\rm RC},D_{\rm LC}\approx \Delta = 10\,\Omega$, the current start getting suppressed with higher dephasing. {For comparison, we have placed the no-measurement values of the occupation $\rho_{\rm CC}^{(0)}$ and tunneling current $\IT^{(0)}$ on the color bar of respective panels. $\rho_{\rm CC}$ improves over this value in the whole parameter space, while $\IT$ in almost the whole space except when $D_{\rm LC}\to0$ and $D_{\rm RC}$ is finite. The reduction of the current shown in Fig.~\ref{fig:steady_state}(b) for $\gL=\gC=0$ and $\gR\neq0$ is hence extremely fragile upon detecting the electron in the nearby dots. }

Our description allows us to analytically estimate the condition when $\IT$ is maximized, assuming that $\Omega$ is much smaller than the other energy scales. The leading term of $\IT$ in an expansion in $\Omega$:
\begin{equation}
I_{\rm T} \approx \frac{2D_{\rm LC}\,A\,\Omega^2}{D_{\rm LC}\,A\,(A+D_{\rm LC}) + \Delta^2(A+4D_{\rm LC})},
\end{equation}
where $A=2D_{\rm RC} + \GR$, is maximized for
\begin{equation}\label{eq:optimal_current}
    D_{\rm LC}=\Delta \text{ and } D_{\rm RC}=\Delta-\frac{\Gamma_\Rb}{2}, 
\end{equation}
given that $2\Delta>\GR$. 
This condition is marked with a star in Fig.~\ref{fig:ss_colormap}(b).
In terms of the measurement strengths, this can be achieved with $\gamma_\Ld=0$, $\gamma_\Cd=20\,\Omega$, and $\gamma_\Rd = 1.02\,\Omega$. Thus, we can get close to the optimal current by just measuring the central dot with the appropriate strength of $2\,\Delta$, since the current varies smoothly in this parameter regime.

Now, we analyze the region of measurement-assisted tunneling and the optimal current as a function of barrier height $\Delta$. In Fig.~\ref{fig:IT_over_I0}, we plot the TQD current, divided by $I_0$, as a function of the barrier height for different measurement scenarios. The reference value, $I_0$, is the TQD current without measurement, which also varies with $\Delta$. This ratio gives a measure of the current enhancement we get with measurement-assisted tunneling. The black curve shows how the enhancement in optimal tunneling current, calculated using Eq.~\eqref{eq:optimal_current}, varies with $\Delta$ and we observe a linear rise. For comparison, we also show that relaxing the optimal conditions to just having $\gamma_{\rm C}=\gamma^*=2\Delta$ allows for a close-to-optimal current, see the orange curve in Fig.~\ref{fig:IT_over_I0}. 
It is important to note that even though the measurement-assisted tunneling improves with $\Delta$ in a relative sense, the optimal current does actually decrease with $\Delta$, as shown in the inset.

The other two cases shown have a fixed $\gC$, where we find that the
steady-state current is close to the optimal value for $\Delta\lesssim\gC$, but then deviates from it for higher $\Delta$ (see Fig.~\ref{fig:IT_over_I0}). These results suggest that near-optimal current enhancement occurs when the measurement strength at the barrier is comparable or higher than the barrier height.

\section{Conclusion}
\label{sec:conclusion}

We investigated nonequilibrium transport in a TQD system under a global continuous-measurement scheme in which a single QPC monitors multiple dot occupations with independently tunable strengths. Extending the framework of Ref.~\onlinecite{singh2025capturing}, we showed that once monitoring becomes spatially structured, the steady state is not governed by the absolute measurement strength, but by the relative dephasing rates induced across different site pairs. 
{For local detections in which only a single dot is measured, we find contrasting behaviors depending on which dot is being measured. In the strong measurement limit, finite barrier occupation can only be achieved when the barrier C itself is being measured. The tunneling current vanishes asymptotically in all cases because of the quantum Zeno effect. In the intermediate regime, measuring either dot L or C results in high barrier occupation and tunneling current. However, measuring R does not meaningfully affect barrier occupation and only suppresses tunneling current.}

By engineering relative measurement strengths, one can further enhance virtual-state occupation in the intermediate and strong regime, and steady-state current over an intermediate regime. 
With strong measurements, the steady state becomes independent of the underlying Hamiltonian parameters, and the virtual-state occupation approaches $1/2$ when the right bond is dephased strongly but not the left bond. 
{By analytically optimizing the measurement strengths, we show that the current is maximized when the dephasing rates satisfy $D_{\rm LC} = \Delta$ and $D_{\rm RC} = \Delta - \Gamma_{\rm R}/2$. This optimal configuration enhances transport by increasing the occupation of the virtual state while avoiding Zeno suppression. Notably, we find that near-optimal current can be achieved with a significantly simpler configuration in which only the central dot is monitored with strength $\gamma_{\rm C} = 2\Delta$.}


Continuous quantum measurement has also recently been explored as a thermodynamic resource, where the information extracted from monitoring can be converted into directed transport or useful work in measurement-powered engines and refrigerators~\cite{elouardefficient2018,bresque_twoqubit_2021,ferreira_transport_2024,bhandari2023measurement,jordanquantum2020,gherardini2020stabilizing,mitchison2021charging,mohammady2017quantum,yanik2022thermodynamics,trigal2026noninvasive,litzba_coupling_2026}. In Ref.~\onlinecite{sanchez2026making}, we examined this idea in the context of a single-dot detection scheme, showing how measurement backaction can drive heat-engine functionality, autonomous refrigeration and state purification.
The results presented in this article suggest that global monitoring could further improve the performance of these thermodynamic operations beyond the single-dot detection scheme. Since relative dephasing rates provide additional control over steady-state populations and currents, they may enable enhanced energy exchange and transport properties in measurement-driven devices.



\section*{Acknowledgments}
We thank Natalia Ares and John Nichol for valuable discussions. A.N.S. and A.N.J. acknowledge support from the John Templeton Foundation Grant ID 63209. R.S. acknowledges funding from the Spanish Ministerio de Ciencia e Innovaci\'on via grant No. PID2024-157821NB-I00, and through the ``Mar\'{i}a de Maeztu'' Programme for Units of Excellence in R{\&}D CEX2023-001316-M.

\appendix

\section{Weak Measurement Condition}\label{app:diffusive_condition}
For individual measurement readouts to only weakly impact the quantum trajectories, the deviations in QPC's transmittance should be much smaller than its average value~\cite{andrew_book}. The transmittance is given by ${\cal T}_{\rm QPC}=(1+e^{-\epsilon})^{-1}$, where $\epsilon = 2\pi(E - \tilde{V} -  \hbar\omega_y/2)/\hbar \omega_x$, $\omega_x$ and $\omega_y$ characterize the saddle potential, $E$ is the energy of the scattering electron, and $\tilde{V}$ is the potential induced by the TQD electron~\cite{buttiker_quantized_1990}. Now, suppose that when the TQD electron is fully occupied in $i \in$~L/C/R, then $\tilde{V}$ is given by $ V_i$ respectively. If the electron is in a superposition state $\alpha \ket{\rm L} + \beta\ket{\rm C} + \gamma\ket{\rm R}$, $\tilde{V}$ is given by $\alpha  V_{\rm L} + \beta V_{\rm C} + \gamma V_{\rm R}$. It is clear to see that the condition, $|{\cal T}_i-{\cal T}_0|\ll {\cal T} = ({\cal T}_0+\Sigma_i {\cal T}_i)/4$, is sufficient to ensure individual readouts are weak.

\section{Stochastic Master Equations}\label{app:SME}

The quantum state evolution can be evaluated using the stochastic master equation in Eq.~\eqref{eq:ME} together with the explicit forms of the bath and measurement superoperators. In the global measurement configuration considered here, the QPC monitors all three dot occupations with independently tunable strengths. Because the measurement operator is diagonal in the local charge basis, it does not directly induce transitions between sites; instead, it suppresses coherences between states at rates determined by differences in the corresponding measurement strengths.

Using the jump operator defined in Eq.~\eqref{eq:meas_jump_op},
the measurement superoperator $\mathcal{L}_\gamma \hat{\rho}$ from Eq.~\eqref{eq:measurement_superoperator} can be written explicitly as
\begin{widetext}
\begin{align}
\mathcal{L}_\gamma \hat{\rho} = \ &
\bigg[\gamma_\Ld \rho_{\Ld\Ld} \ket{\rm L}\bra{\rm L} + \gamma_\Cd \rho_{\Cd\Cd} \ket{\rm C}\bra{\rm C} + \gamma_\Rd \rho_{\Rd\Rd} \ket{\rm R}\bra{\rm R} \nonumber  + \sqrt{\gamma_\Ld \gamma_\Cd} \left( \rho_{\Ld\Cd} \ket{\rm L}\bra{\rm C} + \rho_{CL} \ket{\rm C}\bra{\rm L} \right) \\
& + \sqrt{\gamma_\Cd \gamma_\Rd} \left( \rho_{CR} \ket{\rm C}\bra{\rm R} + \rho_{\Rd\Cd} \ket{\rm R}\bra{\rm C} \right)
 + \sqrt{\gamma_\Ld \gamma_\Rd} \left( \rho_{\Ld\Rd} \ket{\rm L}\bra{\rm R} + \rho_{RL} \ket{\rm R}\bra{\rm L} \right)\bigg] \nonumber \\
& - \frac{1}{2} \bigg[ \left( \gamma_\Ld \ket{\rm L}\bra{\rm L} + \gamma_\Cd \ket{\rm C}\bra{\rm C} + \gamma_\Rd \ket{\rm R}\bra{\rm R} \right) \hat{\rho}
+ \hat{\rho} \left( \gamma_\Ld \ket{\rm L}\bra{\rm L} + \gamma_\Cd \ket{\rm C}\bra{\rm C} + \gamma_\Rd \ket{\rm R}\bra{\rm R} \right) \bigg]  \\
& + \frac{dW}{dt} \Bigg[
\left( \sqrt{\gamma_\Ld} \ket{\rm L}\bra{\rm L} + \sqrt{\gamma_\Cd} \ket{\rm C}\bra{\rm C} + \sqrt{\gamma_\Rd} \ket{\rm R}\bra{\rm R} \right) \hat{\rho}
+ \hat{\rho} \left( \sqrt{\gamma_\Ld} \ket{\rm L}\bra{\rm L} + \sqrt{\gamma_\Cd} \ket{\rm C}\bra{\rm C} + \sqrt{\gamma_\Rd} \ket{\rm R}\bra{\rm R} \right) \nonumber \\
&\qquad\qquad - 2 \left( \sqrt{\gamma_\Ld} \rho_{\Ld\Ld} + \sqrt{\gamma_\Cd} \rho_{\Cd\Cd} + \sqrt{\gamma_\Rd} \rho_{\Rd\Rd} \right) \hat{\rho} \Bigg].\nonumber
\end{align}
\end{widetext}

The first group of terms above corresponds to the usual Lindblad dissipator associated with continuous monitoring. These deterministic contributions generate dephasing between different charge configurations. Here, the off-diagonal terms in the density matrix are a result of a single QPC measuring the occupation of different dots simultaneously. Had we considered three QPCs each measuring individual dot $j$ with strength $\gamma_{j}$, where $j \in {\rm \{L, C, R\}}$, we would have recovered all the terms in the above superoperator except the off-diagonal ones. The term proportional to $dW/dt$ represents the stochastic backaction of the measurement record and differentiates individual quantum trajectories from their ensemble average. Upon averaging over realizations, the stochastic contributions vanish and only the deterministic dephasing remains.

Similarly, the Lindblad superoperators describing the coupling to the thermal reservoirs, taken in the large bias limit, can be written as

\begin{equation}
    \mathcal{L}_{\Lb}\hat{\rho} = \Gamma_{\Lb} \left(\rho_{00}\ket{\rm L}\bra{\rm L} - \frac{1}{2}(\ket{\rm 0}\bra{\rm 0}\hat{\rho} + \hat{\rho}\ket{\rm 0}\bra{\rm 0}) \right),
\end{equation}

\begin{equation}
    \mathcal{L}_{\Rb}\hat{\rho} = \Gamma_{\Rb} \left(\rho_{\Rd\Rd}\ket{\rm 0}\bra{\rm 0} - \frac{1}{2}(\ket{\rm R}\bra{\rm R}\hat{\rho} + \hat{\rho}\ket{\rm R}\bra{\rm R}) \right).
\end{equation}

In this limit, transport is effectively unidirectional: electrons are injected from the left reservoir and removed at the right reservoir. The bath superoperators therefore modify populations and coherences in a way consistent with irreversible particle exchange while preserving the trace.
These superoperators, together with the coherent Hamiltonian evolution, produce a closed set of stochastic differential equations for each density-matrix element. For clarity, we separate population dynamics from the evolution of coherences. The population equations are
\begin{widetext}
\begin{align}
\dot{\rho}_{\Ld\Ld} &= -2\Omega\, \mathrm{Im}\, \rho_{\Ld\Cd} + \Gamma_{\sf L} \rho_{00}
+ 2 \frac{dW}{dt} \rho_{\Ld\Ld} \left( \sqrt{\gamma_\Ld} - \left( \sqrt{\gamma_\Ld} \rho_{\Ld\Ld} + \sqrt{\gamma_\Cd} \rho_{\Cd\Cd} + \sqrt{\gamma_\Rd} \rho_{\Rd\Rd} \right) \right), \\
\dot{\rho}_{\Cd\Cd} &= 2\Omega \left( \mathrm{Im}\, \rho_{\Ld\Cd} + \mathrm{Im}\, \rho_{\Rd\Cd} \right)
+ 2 \frac{dW}{dt} \rho_{\Cd\Cd} \left( \sqrt{\gamma_\Cd} - \left( \sqrt{\gamma_\Ld} \rho_{\Ld\Ld} + \sqrt{\gamma_\Cd} \rho_{\Cd\Cd} + \sqrt{\gamma_\Rd} \rho_{\Rd\Rd} \right) \right), \\
\dot{\rho}_{\Rd\Rd} &= -2\Omega\, \mathrm{Im}\, \rho_{\Rd\Cd} - \Gamma_{\sf R} \rho_{\Rd\Rd}
+ 2 \frac{dW}{dt} \rho_{\Rd\Rd} \left( \sqrt{\gamma_\Rd} - \left( \sqrt{\gamma_\Ld} \rho_{\Ld\Ld} + \sqrt{\gamma_\Cd} \rho_{\Cd\Cd} + \sqrt{\gamma_\Rd} \rho_{\Rd\Rd} \right) \right).
\end{align}
\end{widetext}
The stochastic contributions in these equations depend on the deviation of each occupation from the measurement-conditioned expectation value. This structure ensures that the ensemble-averaged dynamics reduce to deterministic rate equations while individual trajectories retain measurement-induced fluctuations.

The evolution of the coherences reflects the interplay between coherent tunneling, energy detuning, reservoir-induced damping, and measurement-induced dephasing. For the left–center and right–center coherences, we obtain
\begin{widetext}
\begin{align}
\mathrm{Re}\, \dot{\rho}_{\Ld\Cd} &= -\Delta\, \mathrm{Im}\, \rho_{\Ld\Cd} - \Omega\, \mathrm{Im}\, \rho_{\Ld\Rd} \nonumber \\
&\quad - \left[ \frac{(\sqrt{\gamma_\Ld} - \sqrt{\gamma_\Cd})^2}{2} \mathrm{Re}\, \rho_{\Ld\Cd}
+ 2\frac{dW}{dt} \mathrm{Re}\, \rho_{\Ld\Cd} \left( \sqrt{\gamma_\Ld}(\rho_{\Ld\Ld} - \tfrac{1}{2})
+ \sqrt{\gamma_\Cd}(\rho_{\Cd\Cd} - \tfrac{1}{2}) + \sqrt{\gamma_\Rd} \rho_{\Rd\Rd} \right) \right], \\[6pt]
\mathrm{Re}\, \dot{\rho}_{\Rd\Cd} &= -\Delta\, \mathrm{Im}\, \rho_{\Rd\Cd} - \frac{\Gamma_\Rd}{2} \mathrm{Re}\, \rho_{\Rd\Cd}
+ \Omega\, \mathrm{Im}\, \rho_{\Ld\Rd} \nonumber \\
&\quad - \left[ \frac{(\sqrt{\gamma_\Rd} - \sqrt{\gamma_\Cd})^2}{2} \mathrm{Re}\, \rho_{\Rd\Cd}
+ 2\frac{dW}{dt} \mathrm{Re}\, \rho_{\Rd\Cd} \left( \sqrt{\gamma_\Rd}(\rho_{\Rd\Rd} - \tfrac{1}{2})
+ \sqrt{\gamma_\Cd}(\rho_{\Cd\Cd} - \tfrac{1}{2}) + \sqrt{\gamma_\Ld} \rho_{\Ld\Ld} \right) \right],\\
%
\mathrm{Im}\, \dot{\rho}_{\Ld\Cd} &= -\Omega(\rho_{\Cd\Cd} - \rho_{\Ld\Ld}) + \Delta\, \mathrm{Re}\, \rho_{\Ld\Cd} + \Omega\, \mathrm{Re}\, \rho_{\Ld\Rd} \nonumber \\
&\quad - \left[ \frac{(\sqrt{\gamma_\Ld} - \sqrt{\gamma_\Cd})^2}{2} \mathrm{Im}\, \rho_{\Ld\Cd}
+ 2\frac{dW}{dt} \mathrm{Im}\, \rho_{\Ld\Cd} \left( \sqrt{\gamma_\Ld}(\rho_{\Ld\Ld} - \tfrac{1}{2})
+ \sqrt{\gamma_\Cd}(\rho_{\Cd\Cd} - \tfrac{1}{2}) + \sqrt{\gamma_\Rd} \rho_{\Rd\Rd} \right) \right], \\[6pt]
\mathrm{Im}\, \dot{\rho}_{\Rd\Cd} &= -\Omega(\rho_{\Cd\Cd} - \rho_{\Rd\Rd}) + \Delta\, \mathrm{Re}\, \rho_{\Rd\Cd}
- \frac{\Gamma_\Rd}{2} \mathrm{Im}\, \rho_{\Rd\Cd} + \Omega\, \mathrm{Re}\, \rho_{\Ld\Rd} \nonumber \\
&\quad - \left[ \frac{(\sqrt{\gamma_\Rd} - \sqrt{\gamma_\Cd})^2}{2} \mathrm{Im}\, \rho_{\Rd\Cd}
+ 2\frac{dW}{dt} \mathrm{Im}\, \rho_{\Rd\Cd} \left( \sqrt{\gamma_\Rd}(\rho_{\Rd\Rd} - \tfrac{1}{2})
+ \sqrt{\gamma_\Cd}(\rho_{\Cd\Cd} - \tfrac{1}{2}) + \sqrt{\gamma_\Ld} \rho_{\Ld\Ld} \right) \right],
\end{align}
\end{widetext}
The terms proportional to $\Omega$ and $\Delta$ originate from the Hamiltonian and govern coherent oscillations between sites. The contributions proportional to $(\sqrt{\gamma_i}-\sqrt{\gamma_j})^2$ represent measurement-induced dephasing,
making it clear that equal monitoring of two sites leaves their mutual coherence unaffected {once averaged over stochastic realizations of $dW$.}

Finally, the coherence between the outer dots evolves according to
\begin{widetext}
\begin{align}
\mathrm{Re}\, \dot{\rho}_{\Ld\Rd} &= -\Omega(\mathrm{Im}\, \rho_{\Rd\Cd} + \mathrm{Im}\, \rho_{\Ld\Cd}) - \frac{\Gamma_\Rd}{2} \mathrm{Re}\, \rho_{\Ld\Rd} \nonumber \\
&\quad - \left[ \frac{(\sqrt{\gamma_\Ld} - \sqrt{\gamma_\Rd})^2}{2} \mathrm{Re}\, \rho_{\Ld\Rd}
+ 2\frac{dW}{dt} \mathrm{Re}\, \rho_{\Ld\Rd} \left( \sqrt{\gamma_\Ld}(\rho_{\Ld\Ld} - \tfrac{1}{2})
+ \sqrt{\gamma_\Rd}(\rho_{\Rd\Rd} - \tfrac{1}{2}) + \sqrt{\gamma_\Cd} \rho_{\Cd\Cd} \right) \right], \\[6pt]
\mathrm{Im}\, \dot{\rho}_{\Ld\Rd} &= -\Omega(\mathrm{Re}\, \rho_{\Rd\Cd} - \mathrm{Re}\, \rho_{\Ld\Cd}) - \frac{\Gamma_\Rd}{2} \mathrm{Im}\, \rho_{\Ld\Rd} \nonumber \\
&\quad - \left[ \frac{(\sqrt{\gamma_\Ld} - \sqrt{\gamma_\Rd})^2}{2} \mathrm{Im}\, \rho_{\Ld\Rd}
+ 2\frac{dW}{dt} \mathrm{Im}\, \rho_{\Ld\Rd} \left( \sqrt{\gamma_\Ld}(\rho_{\Ld\Ld} - \tfrac{1}{2})
+ \sqrt{\gamma_\Rd}(\rho_{\Rd\Rd} - \tfrac{1}{2}) + \sqrt{\gamma_\Cd} \rho_{\Cd\Cd} \right) \right].
\end{align}
\end{widetext}
The outer-dots coherence is influenced indirectly by both central-dot tunneling and measurement asymmetry. As in the other coherence equations, the dephasing rate depends only on relative measurement strengths, reinforcing the central role of spatially structured monitoring in shaping the steady state.
The conservation of probability, $\rho_{00} + \rho_{\Ld\Ld} + \rho_{\Cd\Cd} + \rho_{\Rd\Rd} = 1$,
must be imposed in addition to the above equations to numerically evaluate the complete stochastic evolution. Together, these equations fully determine the trajectory-level dynamics of the triple quantum dot under global continuous monitoring.

\vspace{20pt}


\section{Deriving Measurement Strengths}\label{app:measurement_strength}

We now show how the measurement strengths $\gamma_j$ follow from the microscopic QPC current statistics, such that the measurement operator in Eq.~\eqref{eq:meas_jump_op} is recovered.
The QPC current can be written as
\begin{equation}
    I_{\rm Q}(t) = \bar{I}_{\rm Q}(t) + \sqrt{S_I}\,\frac{dW}{dt},
\end{equation}
where $\bar{I}_{\rm Q}(t)$ is the expected current based on the system state, $S_I$ is the shot-noise in current~\cite{blanter_shot_2000},
\begin{equation}
    S_I = (eV)\,\frac{2e^2}{h}\,{\cal T}(1-{\cal T}),
\end{equation}
and $dW$ is a Wiener increment.

If the measurement readout gives a value $I$, the (unnormalized) probability that the system occupies the state $|j\rangle\langle j|$ is Gaussian,
\begin{equation}
    P_j \propto \exp\!\left[-\frac{dt}{2S_I}(I - I_j)^2\right],
\end{equation}
where $I_j$ is the average QPC current associated with state $|j\rangle$.
Substituting the stochastic current $I = \langle I \rangle + \sqrt{S_I}\, dW/dt$, we obtain
\begin{align}
    P_j 
    &\propto \exp\!\left[-\frac{dt}{2S_I}
    \Big(\langle I \rangle + \sqrt{S_I}\frac{dW}{dt} - I_j\Big)^2\right] \\
    &\propto \exp\!\left[-\frac{dt}{2S_I}
    (\langle I \rangle - I_j)^2 
    - \frac{dW}{\sqrt{S_I}}(\langle I \rangle - I_j)\right].
\end{align}
Simplifying it further, we are left with a first-order term in $dW$ and none in $dt$:
\begin{equation}
    P_j \propto 1
    - \frac{dW}{\sqrt{S_I}}(\langle I \rangle - I_j).
\end{equation}
Using Itô calculus, the corresponding expansion of the square root is
\begin{equation}
    \sqrt{P_j} \propto 
    1
    - \frac{(\langle I \rangle - I_j)^2}{8S_I}\,dt
    - \frac{\langle I \rangle - I_j}{2\sqrt{S_I}}\,dW.
\end{equation}
After a current measurement at time $t$, the state at $t+dt$ updates according to
\begin{equation}
    \hat{\rho}(t+dt)
    = \frac{\hat{M}\,\hat{\rho}(t)\,\hat{M}^\dagger}
    {\mathrm{Tr}\!\left(\hat{M}^\dagger \hat{M}\,\hat{\rho}\right)},
\end{equation}
where the Kraus operator is
\begin{equation}
    \hat{M} \propto \sum_j \sqrt{P_j}\,|j\rangle\langle j|.
\end{equation}
All proportionality constants cancel in the normalized update.
Expanding to first order yields the stochastic master equation
\begin{align}
    \frac{d\hat{\rho}}{dt}
    &= - \sum_{jk} 
    \frac{1}{8S_I}
    \big(I_j - I_k\big)^2
    |j\rangle\langle k| \\
    &\quad
    - \sum_{jk}
    \frac{dW/dt}{2\sqrt{S_I}}
    \big(2\langle I \rangle - I_j - I_k\big)
    |j\rangle\langle k|.
\end{align}

This form matches the stochastic master equation derived in Appendix~\ref{app:SME} from the measurement jump operator in Eq.~\eqref{eq:meas_jump_op}, provided the measurement strengths are identified as
\begin{equation}
    \gamma_j
    = \frac{(I_0 - I_j)^2}{4S_I}
    = \frac{eV}{2h}\,
      \frac{({\cal T}_0 - {\cal T}_j)^2}{{\cal T}(1-{\cal T})}.
\end{equation}

\section{Equivalent Measurement Configurations}\label{app:equiv_meas}

From the structure of the dephasing terms, one can see that the jump operators
\begin{equation}
\hat{L}
=
\sqrt{\gL}\,\ket{\mathrm{L}}\bra{\mathrm{L}}
+
\sqrt{\gC}\,\ket{\mathrm{C}}\bra{\mathrm{C}}
+
\sqrt{\gR}\,\ket{\mathrm{R}}\bra{\mathrm{R}},
\end{equation}
and
\begin{equation}
\begin{aligned}
\hat{L}'
&=
\hat{L}
+
c\,(\mathbf{1}-\ket{0}\bra{0}) \\
&=
(\sqrt{\gL}+c)\,\ket{\mathrm{L}}\bra{\mathrm{L}}
+
(\sqrt{\gC}+c)\,\ket{\mathrm{C}}\bra{\mathrm{C}}\\ 
&\qquad +
(\sqrt{\gR}+c)\,\ket{\mathrm{R}}\bra{\mathrm{R}},
\end{aligned}
\end{equation}
lead to the same steady state, where $c$ is a positive real number expressed in the units of $\Omega^{2}$, and \(\mathbf{1}\) is the identity operator.
In order to understand why this happens, consider the corresponding dissipator
\begin{equation}
\mathcal{D}[\hat L]\hat{\rho}
=
\hat L \hat{\rho} \hat L^\dagger
-
\frac{1}{2}\{\hat L^\dagger \hat L,\hat{\rho}\}.
\end{equation}
Writing \(\hat L'=\hat L+c\hat{P}\), with
\begin{equation}
\hat{P}
=
\mathbf{1}-\ket{0}\bra{0}
=
\ket{\mathrm{L}}\bra{\mathrm{L}}
+
\ket{\mathrm{C}}\bra{\mathrm{C}}
+
\ket{\mathrm{R}}\bra{\mathrm{R}},
\end{equation}
one finds
\begin{equation}
\mathcal{D}[\hat L']\hat{\rho}
=
\mathcal{D}[\hat L]\hat{\rho}
+
\mathcal{D}\!\big[c\hat{P}\big]\hat{\rho}
-
c\,\hat{\Pi}_0 \hat{\rho}\hat L^\dagger
-
c\,\hat L \hat{\rho}\hat{\Pi}_0,
\end{equation}
where $\hat \Pi_0=\ket{0}\bra{0}$ and $\hat P=\mathbf{1}-\hat\Pi_0$.

To see the effect of the additional terms, we evaluate their action on the matrix elements.
First, for populations in the transport subspace ($j\in\{\mathrm{L,C,R}\}$),
\begin{equation}
\bra{j}\hat{\Pi}_0 \hat{\rho}\hat L^\dagger\ket{j}
=
\bra{j}\hat L \hat{\rho}\hat{\Pi}_0\ket{j}
=0,
\end{equation}
since $\hat{\Pi}_0\ket{j}=0$. Moreover, $\bra{j}\mathcal{D}[c\hat P]\hat\rho\ket{j}=0$, 
because $\hat P\ket{j}=\ket{j}$ and $\hat P^2=\hat P$. Therefore,
\begin{equation}
\frac{d}{dt}\rho_{jj}\Big|_{\mathcal{D}[\hat L']}
=
\frac{d}{dt}\rho_{jj}\Big|_{\mathcal{D}[\hat L]},
\end{equation}
i.e., the populations are unchanged.

Now, consider coherences involving the empty state. For $j\in\{\mathrm{L,C,R}\}$, $\bra{0}\mathcal{D}[c\hat P]\hat\rho\ket{j}
=
-\frac{c^2}{2}\rho_{0j}$,
and $
\bra{0}\,\hat L \hat{\rho}\hat{\Pi}_0\ket{j}
=0$,
while $\bra{0}\,\hat{\Pi}_0 \hat{\rho}\hat L^\dagger\ket{j}
=
\,\rho_{0j}\sqrt{\gamma_j}$.
Combining the different contributions, we get
\begin{equation}
\frac{d}{dt}\rho_{0j}\Big|_{\mathcal{D}[\hat L']}
=
\frac{d}{dt}\rho_{0j}\Big|_{\mathcal{D}[\hat L]}
-
\left(\frac{c^2}{2}+c\sqrt{\gamma_j}\right)\rho_{0j}.
\end{equation}

The rest of the coherence terms are not affected by the transformation. Thus, the only direct effect of the transformation $\hat L\to \hat L+c\hat P$ is on the coherences $\rho_{0j}$.

However, in the original master equation, the variables $\rho_{0j}$ are already decoupled from the closed set of equations governing the populations $\rho_{00},\rho_{\rm LL},\rho_{\rm CC},\rho_{\rm RR}$ and the inter-dot coherences $\rho_{\rm LC},\rho_{\rm RC},\rho_{\rm LR}$. The transformation does not change this decoupling structure: it only modifies the equations within the $\rho_{0j}$ sector itself. Therefore, the steady-state solution of the transport sector is unchanged.
Since all steady-state observables discussed in the main text are determined entirely by this closed sector, the part of the stationary density matrix relevant for transport is identical for $\hat L$ and $\hat L'$.


\section{Uniform Detection}\label{app:equal_measurement}

Here we analyze the special case in which all three measurement strengths are equal,
\begin{equation}
\gamma_{\rm L}=\gamma_{\rm C}=\gamma_{\rm R}.    
\end{equation}

\begin{figure}
    \centering
    \includegraphics[width=\linewidth]{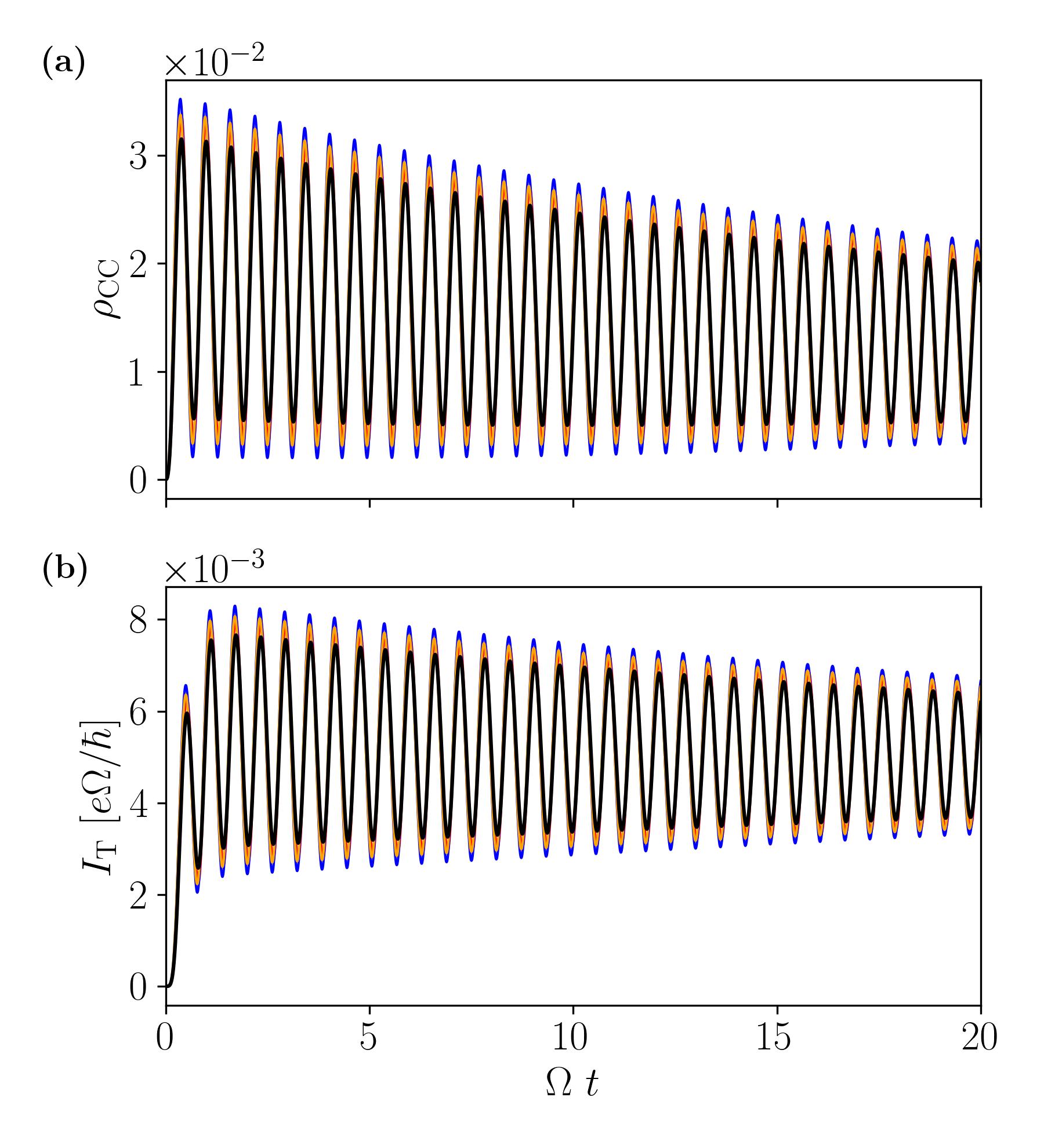}
    \caption{Central dot occupation (a), and TQD current (b) plotted as a function of time. The initial state was taken to be $\ket{\rm 0}\bra{\rm 0}$. The black curve correspond to the ensemble-averaged evolution, and the colored curves are individual stochastic trajectories. For the latter, time averaging is done over a rectangular window of $0.1~\Omega^{-1}$. Parameters: $\Gamma_{\Lb} = 10~\Omega$, $\Gamma_\Rb = 8~\Omega$, $\Delta=10~\Omega$, and $\gamma_\Ld = \gamma_\Cd = \gamma_\Rd = \Omega$.
    }
    \label{fig:app_stoch_traj}
\end{figure}

In this limit, the measurement operator becomes proportional to the total occupation operator of the single-electron subspace. The measurement jump operator can be written as
\begin{equation}
L = \sqrt{\gamma}\,(\mathbf{1}-|0\rangle\langle 0|),
\end{equation}
which projects only between the empty state $|0\rangle$ and the one-electron subspace. The TQD Hamiltonian acts entirely within the one-electron subspace, and therefore commutes with this projector. This provides a concrete example of the general invariance discussed in the previous section. In the equal-strength case, the measurement operator reduces to a projector onto the one-electron subspace, which commutes with the Hamiltonian. As a result, the measurement does not affect coherences within this subspace, and the ensemble-average evolution remain independent of $\gamma$, as shown in Fig.~\ref{fig:app_stoch_traj}.

The situation is different at the level of conditional trajectories. Although the unconditional (ensemble-averaged) master equation is unchanged, individual realizations depend on the stochastic measurement record. In the diffusive limit, the Wiener increment $dW$ introduces multiplicative noise terms proportional to the occupation probabilities. These terms modulate the instantaneous amplitudes of the coherences and populations along each trajectory. Consequently, different realizations exhibit different oscillation amplitudes.

However, the phase structure of the oscillations is governed solely by the coherent Hamiltonian dynamics. Since the measurement operator commutes with the Hamiltonian within the one-electron subspace, it does not introduce any relative phase shifts between the dot states. The stochastic terms rescale amplitudes but do not alter the oscillation frequency, which remains set by $\Omega$ and $\Delta$. Ultimately, coupling to the electronic reservoirs drives the system to a steady state. In this equal-strength configuration, the only coherences that survive at long times are $\Im(\rho_{\rm LC})$ and $\Im(\rho_{\rm RC})$, both directly proportional to the transport current. The measurement modifies the conditional fluctuations around this steady state but leaves the ensemble-averaged steady state unchanged.


\section{Steady-state ensemble-averaged occupations for local detection}
\label{app:occupations}
In this section we give the full analytical expressions for the steady-state occupation of the different states resulting from solving Eq.~\eqref{eq:stationarymastereq} when only one of the quantum dots, labeled $l$, is coupled to the detector. In all cases, the denominator ${\cal A}_l$ is defined to fulfill the normalization $\sum_j\rho_{jj}=1$.

\subsection{$D_{\rm LC}=D_{\rm LR}=\gL/2$, $D_{\rm RC}=0$}
In the case where only $\gL$ is finite, the steady state occupations read:
\begin{gather}
\begin{aligned}
\rho_{00}&=4\Gamma_{\sf R}\Omega^2(\Gamma_{\sf R}\gL+4\Omega^2)/{\cal A}_{\rm L}\\
\rho_{\rm LL}&=\Gamma_{\sf L}\{4\Gamma_{\sf R}^2(\Delta^2+\Omega^2)+16\Omega^2\\
&+\Gamma_{\sf R}\gL[\Gamma_{\sf R}(\Gamma_{\sf R}+\gL)+4(\Delta^2+2\Omega^2]\}/{\cal A}_{\rm L}\\
\rho_{\rm CC}&=\Gamma_{\sf L}\{4\Gamma_{\sf R}^2\Omega^2{+}16\Omega^4{+}\Gamma_{\sf R}\gL[\Gamma_{\sf R}^2{+}4(\Delta^2{+}\Omega^2)]\}/{\cal A}_{\rm L}\\
\rho_{\rm RR}&=4\Gamma_{\sf L}\Omega^2(\Gamma_{\sf R}\gL+4\Omega^2)/{\cal A}_{\rm L}.
\end{aligned}
\end{gather}

\subsection{$D_{\rm LC}=D_{\rm RC}=\gC/2$, $D_{\rm LR}=0$}
In the case where $\gC$ is finite, we get:
\begin{gather}
\begin{aligned}
\rho_{00}&=4\Gamma_{\sf R}\Omega^4[4\Gamma_{\sf R}\Omega^4+\Gamma_{\sf R}\gC(\Gamma_{\sf R}\gC)+8\gC\Omega^2]/{\cal A}_{\rm C}\\
\rho_{\rm LL}&=\Gamma_{\sf L}\{(\Gamma_{\sf R}+2\gC)[\Gamma_{\sf R}^2(\gC(\Gamma_{\sf R}+\gR)+4\Delta^2)+16\Omega^4]\\
&+4\Gamma_{\sf R}[5\gC(\Gamma_{\sf R}+\gC)+\Gamma_{\sf R}^2]\Omega^2]\}/{\cal A}_{\rm C}\\
\rho_{\rm CC}&=\Gamma_{\sf L}\{4\Gamma_{\sf R}\Omega^2(\Gamma_{\sf R}^2+\Omega^2)\\
&+\gC[\Gamma_{\sf R}^2((\Gamma_{\sf R}+\gC)^2+4\Delta^2)+4\Gamma_{\sf R}(4\Gamma_{\sf R}+3\gC)\Omega^2\\
&+32\Omega^4]\}/{\cal A}_{\rm C}\\
\rho_{\rm RR}&=4\Gamma_{\sf L}\Omega^4[4\Gamma_{\sf R}\Omega^4+\Gamma_{\sf R}\gC(\Gamma_{\sf R}\gC)+8\gC\Omega^2]/{\cal A}_{\rm C}.
\end{aligned}
\end{gather}

\subsection{$D_{\rm RC}=D_{\rm LR}=\gR/2$, $D_{\rm LC}=0$}
Finally, in the case where $\gR$ is finite, we get:
\begin{gather}
\begin{aligned}
\rho_{00}&=4\Gamma_{\sf R}\Omega^4/{\cal A}_{\rm R}\\
\rho_{\rm LL}&=\Gamma_{\sf L}[\Gamma_{\sf R}(\Gamma_{\sf R}+\gR)(\Delta^2+\Omega^2)+4\Omega^4]/{\cal A}_{\rm R}\\
\rho_{\rm CC}&=\Gamma_{\sf L}[\Gamma_{\sf R}(\Gamma_{\sf R}+\gR)+4\Omega^2]\Omega^2/{\cal A}_{\rm R}\\
\rho_{\rm RR}&=4\Gamma_{\sf L}\Omega^4/{\cal A}_{\rm R}.
\end{aligned}
\end{gather}

\bibliography{references}
\end{document}